\documentclass[conference]{IEEEtran}
\usepackage[pdftex]{graphicx}
\usepackage{array}
\usepackage{float}

\usepackage{ulem}

\usepackage{clrscode}
\usepackage{listings}
\usepackage{fancyvrb}

\usepackage{booktabs}
\usepackage{amsmath}

\usepackage[normal]{caption2}

\ifCLASSINFOpdf
\else
\fi
\begin{document}
\title{Take up DNSSEC When Needed}


%\author{\IEEEauthorblockN{Zheng Wang}
%\IEEEauthorblockA{Qingdao University\\
%Qingdao, Shandong  266071, China\\
%Email: zhengwang09@126.com}}

\author{\IEEEauthorblockN{Zheng Wang}
\IEEEauthorblockA{National Institute of Standards and Technology\\
Gaithersburg, MD 20899, USA\\
zhengwang98@gmail.com}}

%\and
%\IEEEauthorblockN{Homer Simpson}
%\IEEEauthorblockA{Twentieth Century Fox\\
%\and
%\IEEEauthorblockN{James Kirk\\ and Montgomery Scott}
%\IEEEauthorblockA{Starfleet Academy\\
%San Francisco, California 96678-2391\\


%Telephone: (800) 555--1212, Fax: (888) 555--1212}


\maketitle

\begin{abstract}
%\boldmath
The threats of caching poisoning attacks largely stimulate the deployment of DNSSEC. Being a strong but demanding cryptographical defense, DNSSEC has its universal adoption predicted to go through a lengthy transition. Thus the DNSSEC practitioners call for a secure yet lightweight solution to speed up DNSSEC deployment while offering an acceptable DNSSEC-like defense. This paper proposes a new defense against cache poisoning attacks, still using but lightly using DNSSEC. In the solution, DNS operates in the DNSSEC-oblivious mode unless a potential attack is detected and triggers a switch to the DNSSEC-aware mode. The performance of the defense is analyzed and validated. The modeling checking results demonstrate that only a small DNSSEC query load is needed to ensure a small enough cache poisoning success rate.
\end{abstract}


%\keywords{negative caching, NSEC/NSEC3 record, cache hit rate, cache consistency} % NOT required for Proceedings

\section{Introduction}
%\begin{codebox}
%\Procname{$\proc{Insertion-Sort(A)}$}
%\li \For $j \gets 2$ \To $\id{length}[A]$    \label{li:for}
%\li     \Do $\id{key} \gets A[j]$            \label{li:for-begin}
%\li         \Comment Insert $A[j]$ into the sorted sequence $A[1 \twodots j-1]$.
%\li         $i \gets j-1$
%\li         \While $i>0$ and $A[i]>\id{key}$ \label{li:while}
%\li            \Do $A[i+1] \gets A[i]$       \label{li:while-begin}
%\li                $i \gets i-1$             \label{li:while-end}
%                \End
%\li         $A[i+1] \gets \id{key}$          \label{li:for-end}
%        \End
%\end{codebox}

Domain Name System (DNS) is today's largest name resolution system in use. Most of Internet applications rely on DNS to translate human friendly names to addresses, services, servers, etc. However, the early design of DNS did not pay sufficient attention to its security in 1980s. Thus the emerging security problems of DNS drove the community's efforts on developing DNS security mechanisms. One major progress on securing DNS is DNS Security Extensions (DNSSEC)
\cite{DNSSEC} as a set of specifications agreed by IETF in 2005. DNSSEC provides security capabilities by digitally signing 
DNS data using public-key cryptography.
%When a resolver issues a DNS query for a resource record in a DNSSEC-signed zone, the response includes not only the requested record but also the signatures for the record. The validity of the signed record is determined via an authentication chain following the DNS hierarchy.

Largely due to its enormous costs in technology, policy, management, the value of the DNSSEC effort is debatable over years before 2008. Some temporary mitigating solutions were proposed in those years in a hope to exempt the DNS from the heavyweight DNSSEC. However, the discovery of Kaminsky vulnerabilities \cite{Kaminsky} ended the debate since no lightweight mitigating solutions convincingly secure the DNS from the Kaminsky attacks. Until then, the DNS community finally reached a consensus that DNSSEC should roll out as a global Internet infrastructure upgrade.

The Internet-scale DNSSEC deployment requires substantial costs, efforts, coordination, and time. In the measurements of the cost of DNSSEC deployment \cite{Performance}, both cache resolver and authoritative name server are demonstrated to suffer significant performance penalty when turning on DNSSEC. As a large-scale cryptographic system, DNSSEC heavily relies on the secure key generation, storage, distribution and rollover as well as zone signing and record authentication operations. Another important and indispensable aspect that may be underestimated is the changing and increasing operational procedures and policies that may materially affect the life cycle of the DNS data. Hence despite that the DNS community has been sparing no efforts to use its resources to encourage DNS registries, ISPs and enterprises to upgrade to DNSSEC, global DNSSEC deployment is still very far from completion. While the root and most top-level domains have rolled out DNSSEC, the rare adoption on lower level domains, which are mostly the ultimate destination of individual DNS lookups, might shed light on a low level of optimism on the universal DNSSEC adoption as a whole. In some ways, the progress of DNSSEC deployment is similar to the undergoing IPv4 to IPv6 transition: slow and incremental.

The growing threats to DNS in recent years propel the efforts of speeding up DNSSEC deployment. This is because DNSSEC deployment is much like an ``all or nothing'' proposition. In other words, incomplete or halfway DNSSEC deployment is likely to leave a much larger subset of the entire domain name space vulnerable than we may expect. The strong defense against Kaminsky attacks provided by DNSSEC virtually takes effects only when DNSSEC is fully deployed across the Internet -- from the DNS root zone at the top of the DNS hierarchy down to individual top-level domains (such as .com and .net), second-level domains, lower level domains, and even leaf domains in the DNS tree. Not only the target domain itself but also all its ancestor domains (including the parent domain) must be signed to ensure a complete trust chain to get protected by DNSSEC. Otherwise if only any domain in the trust chain turns DNSSEC oblivious, the target domain turns vulnerable to Kaminsky attacks.

To shorten the transition to DNSSEC, one effective way is to lower the costs of DNSSEC while retaining the comparable security capability of DNSSEC. The transition solution should also be able to readlily upgrade to the full DNSSEC whenever necessary or the transition is to be finished.
%The strong defense against Kaminsky attacks provided by DNSSEC virtually takes effects only when DNSSEC is fully deployed across the Internet -- from the DNS root zone at the top of the DNS hierarchy down to individual top-level domains (such as .com and .net), second-level domains, lower level domains, and even leaf domains in the DNS tree. Not only the target domain itself but also all its ancestor domains (including the parent domain) must be signed to ensure a complete trust chain to get protected by DNSSEC. Otherwise if only any domain in the trust chain turns DNSSEC oblivious, the target domain turns vulnerable to Kaminsky attacks. So the DNSSEC deployment is much like an ``all or nothing'' proposition. That is, incomplete or halfway DNSSEC deployment is likely to leave a much larger subset of the entire domain name space vulnerable than we may expect.
This paper proposes TDWN (Take up DNSSEC When Needed), a lightweight defense against cache poisoning attacks. TDWN changes DNSSEC operation from the model of persistent-defense to a model of detect-and-defense. Hence DNSSEC can be expected to be not so aggressively used and the vast costs wasted by DNSSEC in the absence of attacks are saved. TDWN makes full use of the detection capability of recursive resolvers to take up DNSSEC whenever needed so that clients can be kept transparent to the defense. Because of its efficiency and efficacy, TDWN can serve as an interim or transition mechanism for spreading and speeding DNSSEC adoption over a long-term transition.

\section{Related Work}

As a non-DNSSEC solution to the DNS security, Fan et al. \cite{Proxy} proposed prevention schemes embedded in so-called security proxy. But their deployment costs are fairly high because security proxies need to be deployed at both authoritative servers and recursive resolvers to support packing and unpacking of all DNS packets with security label. Schomp et al. \cite{Hotnet} proposed a radical change to the existing DNS eco-system for tackling vulnerabilities in shared DNS resolvers: removing shared DNS resolvers entirely and leaving recursive resolution to the clients. However, each individual client conducting its own resolutions may be still targeted by cache poisoning attackers. The solution also has a large performance penalty because the wide-area DNS traffic and DNS server load will grow significantly due to the absence of cache sharing. Sun et al. \cite{DepenDNS} proposed DepenDNS as a countermeasure which query multiple resolvers concurrently to verify a trustworthy answer. The reliability and availability of history response data used by DepenDNS is a great concern. Besides, the performance concern about DepenDNS is when the queries are multiplied, their processing overheads will also be multiplied. Shulman et al. \cite{Shulman} performed a critical study of the prominent defense mechanisms against poisoning attacks by off-path adversaries, concluding that existing `easy-to-deploy' defenses are not so reliable and thus transition to DNSSEC deserves the efforts. Our proposal has an advantage of providing a strong defense comparable to DNSSEC while lowering the cost to a low as some challenge-response defenses.

\section{The Proposed Defense}

To ``condense'' DNSSEC as best as possible while retaining its security capability against cache poisoning attacks, we propose that DNSSEC can coalesce with attack detection to lower its overheads.

\begin{figure}[!t]
\centering
\includegraphics[width=0.9\linewidth]{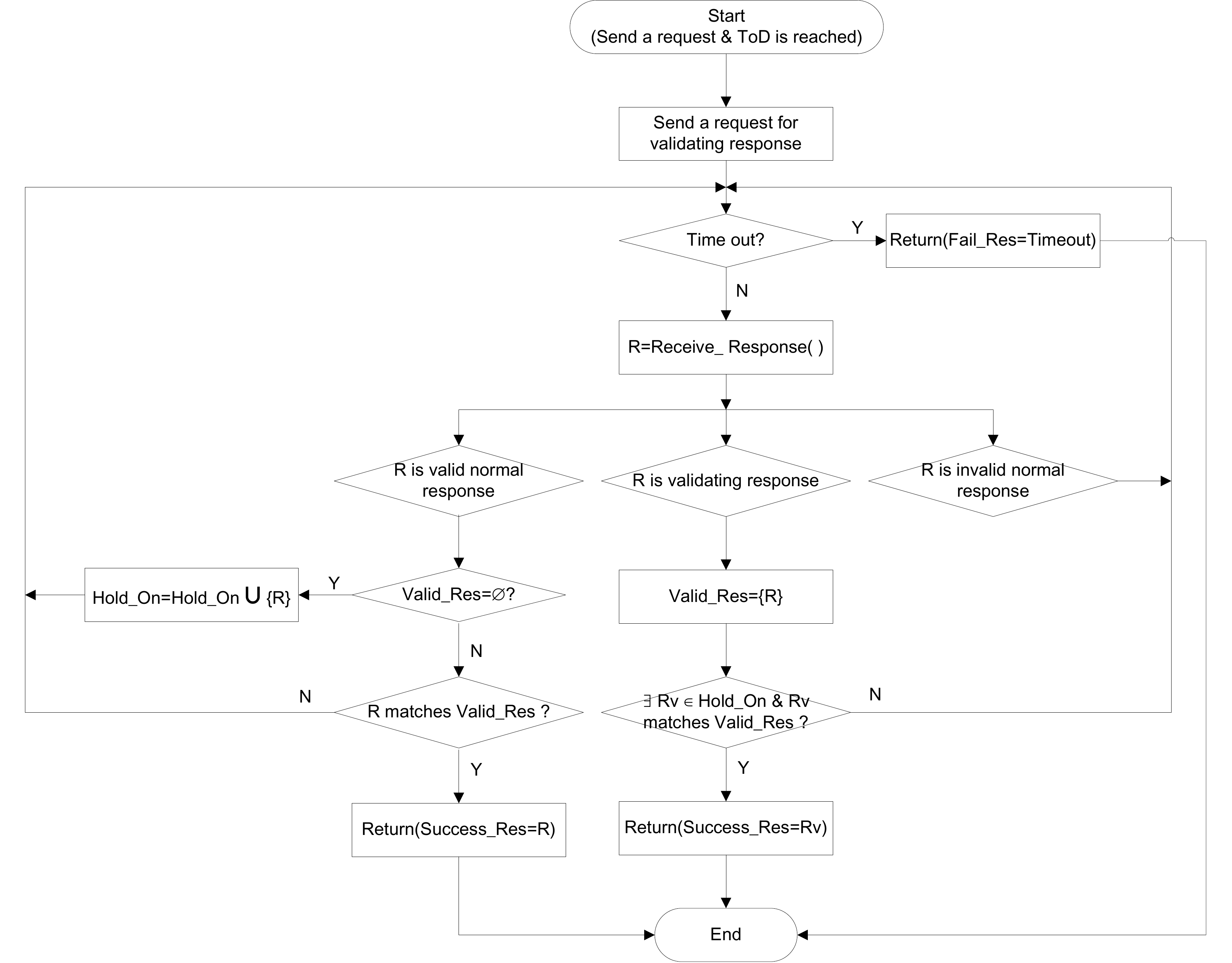}
%\footnotesize{Fig. 1. NSEC record chain}
\caption{DNSSEC-aware mode.}
\vspace{-2em}
\end{figure}

\subsection{Attack Detection}

Cache poisoning is where the attacker manages to inject bogus data into a resolver's cache with carefully crafted and timed DNS packets. A cache poisoned resolver will response with its wrongfully accepted and cached data, make its clients contact the wrong and possibly malicious servers. For the sake of being accepted by the target resolver, cache poisoning attackers have to guess the transaction ID, port number, and source address of the genuine response in their bogus responses. A number of bogus responses of wrong guessing are expected to be found by the target resolver before one bogus response may accidentally succeed. So failure response counting can be utilized to detect possible cache poisoning attacks. For one DNS question, a failure response is defined as the response mismatching the combination of transaction ID, port number, and source address against its outstanding (wait-for-response) request. In particular, a failure response attempt satisfies: \textbf{a)}: It matches the DNS question (or precisely the triple $<qname,qtype,qclass>$) of the outstanding queri(es). Note that attackers may exploit multiple outstanding queries for the same question to significantly increase the success rate of caching poisoning. This is referred to as ``birthday attack''. In that case, more than one outstanding queries may share one question. \textbf{b)}: If a) holds, it mismatches at least one item among transaction ID, port number, and source address of the outstanding queri(es).

As a means of attack detection, the resolver counts the incoming failure responses for the DNS question of the outstanding queri(es) until the count amounts to a threshold of defense (ToD). Then the alert and the corresponding response to attack is triggered. The appropriate setting of ToD should consider: on one hand, a too large value will result in a non-negligible increase of cache poisoning success rate ahead of any defense in place, e.g., the number of forgery responses is in the order of ten thousands to ensure a 50\% chance of compromise in most cases of DNS operations cite{Z1},cite{Z2}; on the other hand, a too small value will too readily trigger the defense. Problem of false positive stands here when non-malicious but negligent users may unintentionally create a small amount of malformed responses which are identified as failure responses. Another exploit of a small threshold is that adversaries may intentionally feed a few failure responses on the target resolver in a bid to overload it with excessive defenses.

\subsection{The Two Modes}

\subsubsection{DNSSEC-Oblivious Mode}

In the DNSSEC-oblivious mode, the recursive resolver operates in compliant with the conventional DNS. That is, it never sends DNSSEC requests or authenticate DNSSEC response unless it is explicitly required by the client (which sets the DO bit in the request). In handling a response, the recursive resolver simply checks and accepts that response if it successfully matches the outstanding query. In comparison with the conventional DNS, the DNSSEC-oblivious mode supplements the attack detection stated above and thereby the transition to the DNSSEC-aware mode once attack is detected. Therefore the costs of the DNSSEC-oblivious mode are almost as low as the conventional DNS. As long as no attack is detected, the DNSSEC-oblivious mode continues as a normalcy.

\subsubsection{DNSSEC-Aware Mode}

When resolving a DNS question, the resolver transitions from the DNSSEC-oblivious mode to the DNSSEC-aware mode when that question is hit by at least ToD failure responses counted in the attack detection. The DNSSEC-aware mode uses DNSSEC transactions to validate suspicious responses to the DNS question potentially targeted by attackers.

The responding process in the DNSSEC-aware mode is illustrated in Fig. 1. When a DNS question is labeled as suspicious by the attack detection, the resolver should immediately initiate a separate DNSSEC-aware request for that question. Once validated by the validators, the response, which is called ``validating response'' hereinafter, is designated as the trustworthy authority for all upcoming responses to that question. Thus all responses arriving prior to the validating response are simply hold on rather than accepted. Note that the hold-on responses may include one bogus response, which may otherwise look as genuine because it completely matches the oustanding query (but, of course, not validated through DNSSEC). The genuine response is also likely to be hold on if it arrives earlier than the validating response.

The validating response, if validated through DNSSEC, is deemed as trustworthy. It is compared against each response in the hold-on list. If there is any hold-on response matching the validating response, the resolution transaction ends with returning the matching response to the client and discarding other hold-on responses if any. Otherwise, if no candidate hold-on response survives the matching check, all existing candidates in hand will be discarded and the resolver will continue to wait for more candidate responses probably still to come until the resolution transaction times out. Then each newly arrived responses, if any, will be likewise checked against the validating response before it can be accepted and returned. In summary, the DNSSEC-aware mode attempts to return the first candidate response matching the validating response until the resolution transaction times out.

\begin{figure}[!t]
\centering
{\includegraphics[width=0.8\linewidth]{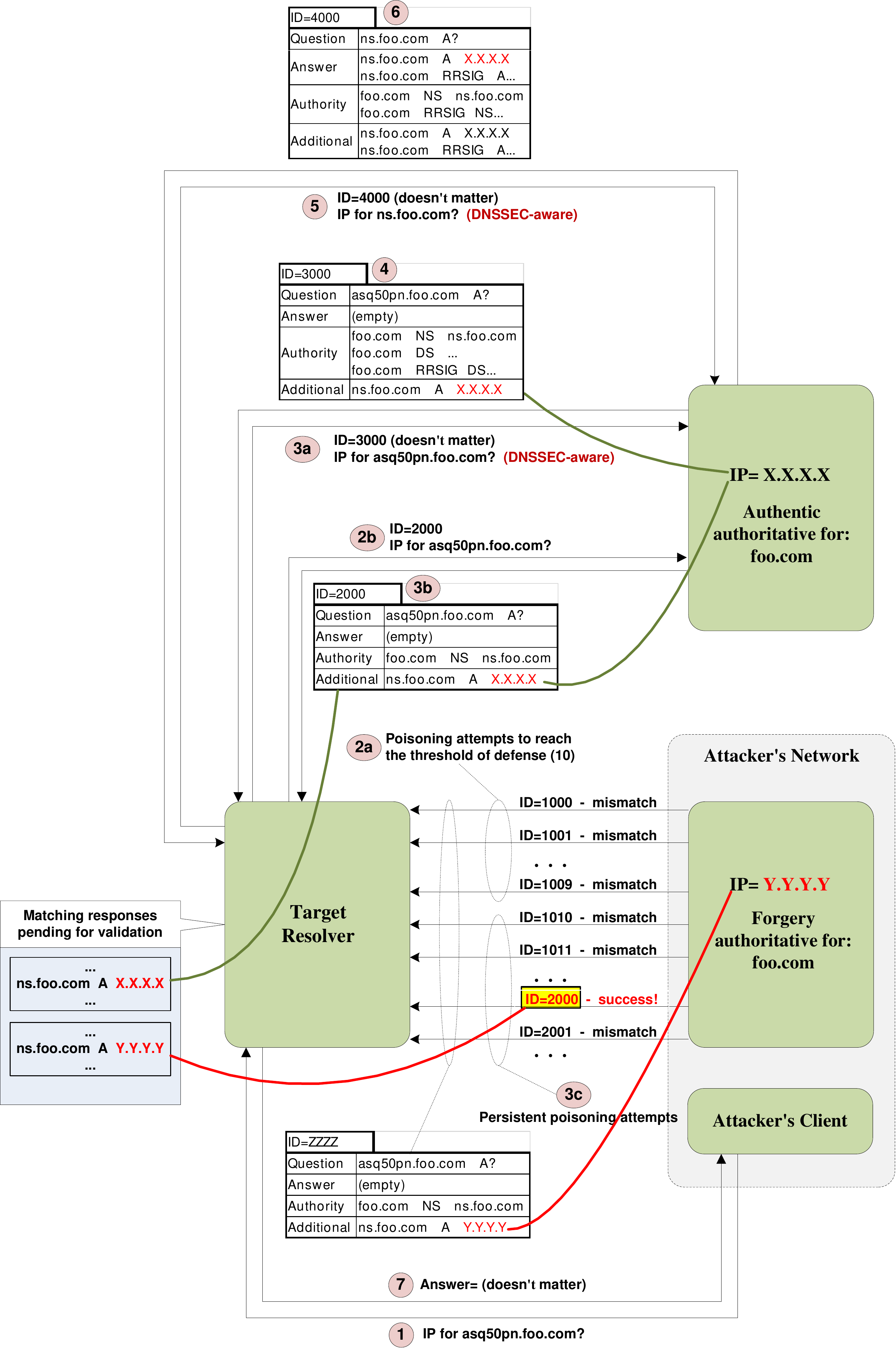}}
%\footnotesize{Fig. 1. NSEC record chain}
\caption{An example of the integration of the two modes.}
\vspace{-2em}
\end{figure}

\subsubsection{Integration of the Two Modes}

We present in detail how the two modes are integrated to defend against cache poisoning attacks. In particular, our example in Fig. 2 shows the defense procedure under the most mighty version of Kaminsky class attacks:
\textbf{(1)}: The attacker's client sends the target resolver a query for the IP address of ``asq50pn.foo.com'' below the target domain ``foo.com''. The domain ``asq50pn.foo.com'' is delicately crafted with random characaters so that it is likely to miss the resolver's cache to trigger an outstanding query.
\textbf{(2a)}: The forgery authoritative server tries to send cache poisoning attempts to the target resolver guessing the transaction ID, etc. of the genius response until the failure responses accumulate to ToD. Each failure response may, e.g., guess a wrong transaction ID, and intends to inject the IP address of the forgery authoritative server, say ``Y.Y.Y.Y''.
\textbf{(2b)}: Roughly in parallel with (2a), the target resolver sends requests to the real authoritative name servers for ``asq50pn.foo.com''.
\textbf{(3a)}: When the attack detection counts the number of failure responses to ToD, the target resolver starts the DNSSEC-aware mode by sending a DNSSEC-aware query for ``asq50pn.foo.com'' soliciting a validating response.
\textbf{(3b)}: Perhaps at the same time as (3a), the genius response arrives at the target resolver informing the IP address of the real authoritative server, say ``X.X.X.X''. However, as the DNSSEC-aware mode is already turned on, the response is hold on rather than simply accepted.
\textbf{(3c)}: The target resolver may still persistently be fed with cache poisoning responses after the ToD failure responses triggers the DNSSEC-aware mode. Before the validation response is returned, the continuous response guessing efforts do have a chance of success. The successful guessing response is also hold on for the future validation.
\textbf{(4)}: When the validating response is obtained by the target resolver, the relevant records in the validating response are subject to DNSSEC validation using the verified public key. That DNSSEC validation may render further DNSSEC transactions such as step (5) and (6) because some signatures (RRSIG records) over the interested data may be absent from the original validating response.
\textbf{(5)}: The target resolver initiates a new DNSSEC transaction to validate the IP address of the authoritative server (``ns.foo.com'').
\textbf{(6)}: The new validating response contains a RRSIG record over the A type (IP address) record of ``ns.foo.com''. By then, the validating response can be validated.
\textbf{(7)}: By checking the hold-on list against the validating response, the IP address of ``ns.foo.com'', namely ``X.X.X.X'', is identified as genius and ``Y.Y.Y.Y'' as bogus. The validated record can thus be used by the target resolver in the final answer as well as in the cache.

\subsection{Aggressive Use of Validating Response}

\begin{figure}[!t]
\centering
{\includegraphics[width=0.9\linewidth]{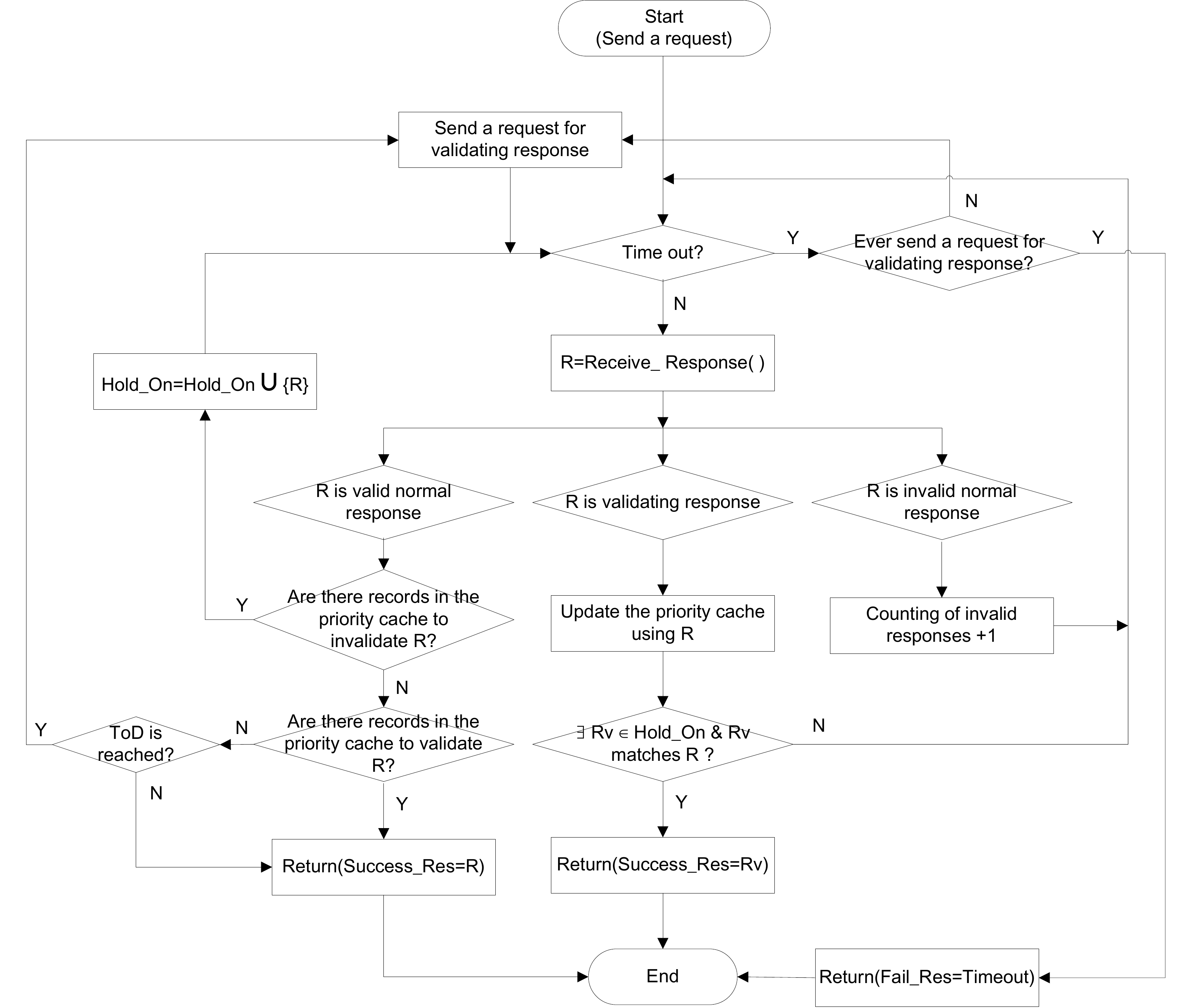}}
%\footnotesize{Fig. 1. NSEC record chain}
\caption{Aggressive use of validating response.}
\vspace{-2em}
\end{figure}

\subsubsection{Efficiency and Security Concerns on the One-Time Use of Validating Response}

As discussed above, the validating response is introduced to defeat the cache poisoning attempt within the resolution transaction of a single query name. However, a validating response is underutilized if it is used only for a single query name, since some data may be shared among different query names. In the example of Fig. 2, the attacker may initiate a query for a new name other than ``asq50pn.foo.com'', say ``b3rr5v.foo.com'', immediately after he receives a genius response (indicating cache poisoning failure) rather than his intended bogus response (indicating cache poisoning success). Because the two query names fall into the same domain ``foo.com'', the data flows of the two defenses almost overlap except for where the query name ``asq50pn.foo.com'' is replaced with ``b3rr5v.foo.com''. But with the one-time use validating response, the resolver need at least two seperate DNSSEC transactions for the two query names respectively. When successive cache poisoning attacks are launched using random generated query names within the same target domain, the resolver would waste a great number of DNSSEC transactions on the largely overlapping data. So the one-time use of validating response is sub-optimal in terms of efficiency.

Being a reasonably small value, ToD still allows for a minor enough chance of caching poisoning success within one window of the TDWN transaction, since the defense leaves the initial ToD-1 caching poisoning attempts free of being detected and validated. While that threat is negligible for a short window, it may grow to serious when the brute force response guessing attack is rapid and continuous for a long window. This is because the success rate of caching poisoning increase dramatically with the number of cache poisoning attempts. The one-time use validating response only defeats the ToD th caching poisoning attempt and its successors within one TDWN transaction, but it virtually does nothing to defend against the initial ToD-1 caching poisoning attempts in the next TDWN transaction. So the one-time use of validating response weakens the TDWN in terms of security.

Based on the above analysis, the underutilized validating response raises not only efficiency concerns but also security concerns. In order to maximize the utilization of validating response and minimize the cache poisoning opportunities, we propose to retain validating responses in cache for a long-lived defense rather than just use them once.

\subsubsection{Caching of Validating Response}

The signed records contained in the validating responses and validated by the recursive resolver should be regarded as more trustworthy than the unsigned records in the valid normal responses. Similar to the conventional DNS caching, those records are cached by the recursive resolver for a period to validate the normal responses. Nevertheless, the caching of validating responses differs from the conventional DNS caching in the following:
\textbf{a)} The validating records are given a priority over the unsigned normal records and they are stored in a priority cache other than a normal cache. Here ``priority'' means: a record in the priority cache can overwrite its unsigned counterparts in the normal cache if they conflict with each other; a record in the priority cache cannot be overwritten by any unsigned record in the more recent normal response; the life cycle of any record in the priority cache is ended either with its TTL expiration or with a replacement by a more recent validating response.
\textbf{b)} The records in the priority cache are basically used for validating normal responses. When a normal response arrives with any record conflicting with the priority cache, the recursive resolver should not accept the response. Instead it waits for its possible successor consistent with the priority cache until the resolution times out. The mechanism of waiting for genius response and denying bogus responses, used by the caching of validating responses, is very similar to that used by the fresh validating responses stated in the DNSSEC-aware mode. But for the sake of maintaining strong priority cache consistency, the recursive resolver should do more than simply return a timeout error as a response in case of resolution timeout.

\subsubsection{Proactive Updating of Validating Response}

One common concern on the Time-to-Live (TTL) based caching such as DNS caching is the weak cache consistency. In DNS caching, a resolver stores a record in the cache as long as specified in that record's TTL field. The typical setting of TTL in DNS records ranges from 1 hour to 1 day. So the change of DNS records in authoritative servers is usually unlikely to be rapidly synchronized to resolvers because the resolvers follow the TTL expiration rule to invalidate the out-of-date cache entries and fetch the up-to-date copies upon requests. In conventional DNS specifications, cache inconsistency only poses a threat to the availability of Internet services because during the cache inconsistency period, the client served with out-to-date DNS records cannot reach the appropriate Internet servers. In the aggressive caching use of validating response, cache inconsistency, however, may result in a serious false positive of genuine response. This is simply because the out-to-date validating records in cache can deny genuine response containing more up-to-date copies of records. When both the genuine response and the bogus responses are invalidated by the stale validating records in cache, a resolution timeout takes place. So a resolution timeout may imply the possibility of cache inconsistency of validating records (and, of course, the possibility of authoritative server unresponsiveness or packet loss in the network).

Thus the hold-on mechanism specified in the DNSSEC-aware mode is slightly changed for caching of validating response. That is, the responses inconsistent with the responses in the priority cache are temporally hold on rather than discarded. Because the inconsistent responses may include the genuine response and the bogus responses in case of cache inconsistency of validating records, they are reserved for further validation.

To still obtain an up-to-date copy of validating record in cache when a resolution timeout (indicating the possibility of cache inconsistency), the resolver should proactively update the validating record in cache by acquiring a fresh validating response. The new validating response will has two usages: validating the hold-on responses and then returning the validated response if any; updating the corresponding validating records in cache. The responding process for the aggressive use of validating response is detailed in Fig. 3.

\section{Performance Analysis}

\begin{figure}[!t]
\centering
\scriptsize{
\begin{Verbatim}[frame=single]
% The present time is initialized at an instance of
% update-triggered query
t=0;  % The time is initialized as zero
T=TTL; % The residual TTL is a full TTL after an
       % update-triggered query
While (){
   if (T==0){ % The residual TTL decreases to zero
       QUERY; % TTL-triggered query
       T=TTL;}
   elseif (UPDATE(t)==true){% An authoritative record
                            % update takes place at t
       QUERY; % Update-triggered query
       T=TTL;}
   t=ELAPSE(t); % Time elapses
   T=T-(ELAPSE(T)-T);} % The residual TTL decreases as time
                       % elapses
\end{Verbatim}
}
\vspace{-1em}
%\footnotesize{Fig. 1. NSEC record chain}
\caption{Process of DNSSEC query event.}
\vspace{-2em}
\end{figure}

%\begin{Verbatim}[frame=single]
%% The present time is initialized at an instance of update-triggered query
%t=0;  % The time is initialized as zero
%T=TTL; % The residual TTL is a full TTL after an update-%triggered query
%While (){
%   if (T=0){      % The residual TTL decreases to zero
%       Query();     % TTL-triggered query
%       T=TTL;}
%   elseif (Update(t)=true){% An authoritative record update
%                           % takes place at t
%       Query();     % Update-triggered query
%       T=TTL;}
%   t=Elapse(t);    % Time elapses
%   T=T-(Elapse(T)-T);} % The residual TTL decreases as time elapses
%\end{Verbatim}

%\begin{lstlisting}
%% The present time is initialized at an instance of update-triggered query
%t=0;  % The time is initialized as zero
%T=TTL; % The residual TTL is a full TTL after an update-%triggered query
%While (){
%   if (T=0){      % The residual TTL decreases to zero
%       Query();     % TTL-triggered query
%       T=TTL;}
%   elseif (Update(t)=true){% An authoritative record update
%                           % takes place at t
%       Query();     % Update-triggered query
%       T=TTL;}
%   t=Elapse(t);    % Time elapses
%   T=T-(Elapse(T)-T);} % The residual TTL decreases as time elapses
%\end{lstlisting}

\subsection{Query Load on the Authoritative Server}

TDWN never initiates DNSSEC transactions unless possible cache poisoning attack is detected at the target resolver. Thus for a vast majority of recursive resolvers which are not constantly targeted by cache poisoning adversaries, TDWN is lightweight in the name resolution cost at both recursive resolvers and authoritative servers because DNSSEC is much less used by TDWN than by the existing DNSSEC deployment strategy.

Consider the case of most severe cache poisoning attack targeting the victim resolver. That is, the attacker continuously sends caching poisoning responses at a high rate towards the target resolver. A DNSSEC transaction is generated by the target resolver if and only if: 1) The validated records in the priority cache expire so that an immediate flurry of caching poisoning responses triggers the DNSSEC-aware mode; 2) No validated response is found until timeout because of the updated authoritative records. As DNSSEC is triggered roughly either by the expiration of TTL or by the updated authoritative records, we first investigate the event of queries triggered by them separately. Without loss of generality, we assume the TTL follows a probability distribution function.

If the target record is heavily requested, the times between successive events (queries) can be approximated by the value of TTL at the instances of events. Let the TTLs or the successive inter-event times are independently and identically distributed. So there is a renewal process in operation for the TTL-triggered queries. Assume that the successive times between the updates of authoritative records are independently and identically distributed. So there is also a renewal process in operation for the update-triggered queries.

However, it is not true that the two renewal processes can be supposed to be independent renewal processes in operation simultaneously. No matter how long the TTL elapses, the update-triggered queries take place merely following the inter-update times. This means the renewal process of update-triggered queries is independent of the renewal process of TTL-triggered queries. But the inter-event times of TTL-triggered queries are dependent of those of update-triggered queries. For example, if there is no update between two successive TTL-triggered queries, their inter-time is a TTL; if there is one update between them, the residual TTL is renewed to a full TTL at the instance of update, and so their inter-time is prolonged to be a full TTL plus a residual TTL; if there is more than one updates between them, the residual TTL is renewed more than one times, and their inter-time becomes a full TTL plus more than one residual TTLs. Given the dependence analyzed above, the sequence of events of DNSSEC queries cannot be considered to be formed by superposing the two individual processes. Instead, we depict the process of DNSSEC queries using the code in Fig. 4.

\subsection{Cache Poisoning Success Rate}

In conventional Kaminsky cache poisoning attacks, the attacker can balance between the number of outstanding requests and the number of bogus response attempts at will to achieve maximum efficiency. Because the number of bogus response attempts is limited for cache poisoning attacks if protected by the proposed scheme, the attacker has to create more duplicate requests for the same target domains subject to bogus response attempts in a bid to increase the probability of successful compromise. However, the number of outstanding requests are also bounded by two aspects in practice: 1) The maximum number of outstanding requests is usually set as a default configuration in most widely used authoritative server implementations. Authoritative servers will thereby discard excessive outstanding requests surpassing the configured limit. So any efforts of producing over-the-limit outstanding requests will prove fruitless. 2) The window allowed to persistently elicit outstanding requests may be bounded by the response time. When the resolver begins receiving a response matching an outstanding request in the wait-for-response list, the list will not necessarily be on the rise since then because the responding rate may be not below the request rate. Hence a conservative estimation of the window of outstanding requests is the response time perceived by the target resolver. As an equivalent of the first limit, the window can be converted to the number of outstanding requests if the sending rate is constant. In summary, the maximum number of outstanding requests is the minimum of the two limits stated above.

Let the threshold of bogus response attempts be $H$, and the maximum number of outstanding requests be $D$. We can express the cumulative probability of cache poisoning failure in all attempts up to and including the H th attempt as
\begin{equation}
\begin{aligned}
P_D(H)&=P(the~1st~attempt~misses,~the~2nd~attempt\\
&misses,~...,~the~H~th~attempt~misses~\mid~D\\
&identical~outstanding~queries)
\end{aligned}
\end{equation}
If $H\ll(I+P)*N$, $P_D(H)$ can be written as
\begin{equation}
P_D(H) = (1 - D / ((I+P)*N))^H
\end{equation}
If no window extension mechanism is applied, the window allocated for launching the $H$ th attempts equals the window of eliciting outstanding requests plus the window of validating the responses. As analyzed above, the first is roughly the response time and the second is also approximated as the response time. That is, one round of cache poisoning attempt takes two response times to obtain a success rate of $1-P_D(H)$. The success rate of cache poisoning within $i$ rounds of cache poisoning attempt is $1 - P_D(H)^i$.

\begin{figure}[!t]%{\captionstyle{normal}}
%\begin{minipage}[b]{0.5\linewidth}
\begin{minipage}[t]{0.48\linewidth}
\centering
\includegraphics[scale=0.31]{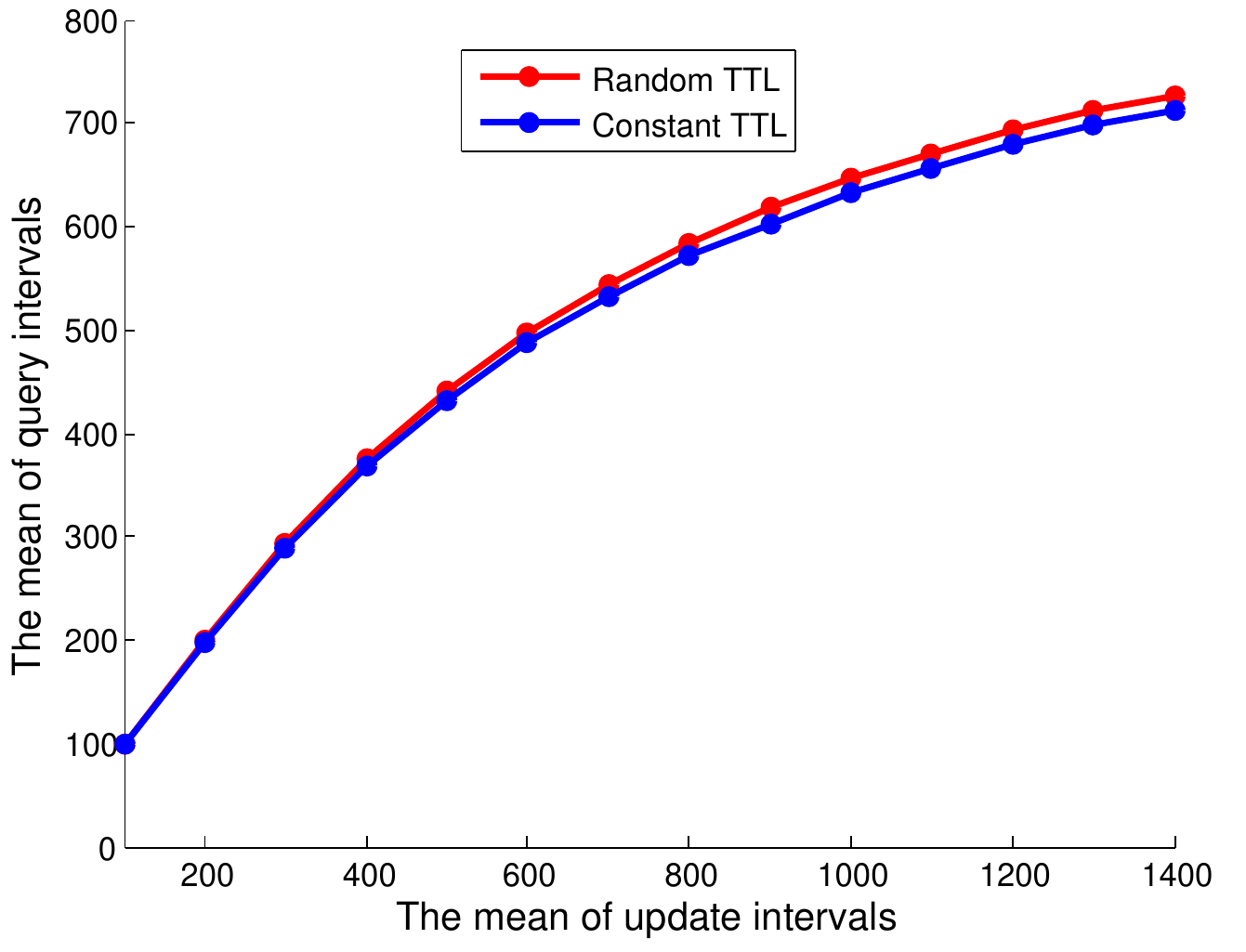}
\vspace{-1em}
%\captionstyle{normal}
\caption{DNSSEC query intervals vs authoritative update intervals.}
%\begin{flushleft}
%\footnotesize{Fig. 1. NSEC record chain}
\vspace{-2em}
%\end{flushleft}
\end{minipage}
\hspace{0.3cm}
%\begin{minipage}[b]{0.5\linewidth}
\begin{minipage}[t]{0.48\linewidth}
\centering
\includegraphics[scale=0.31]{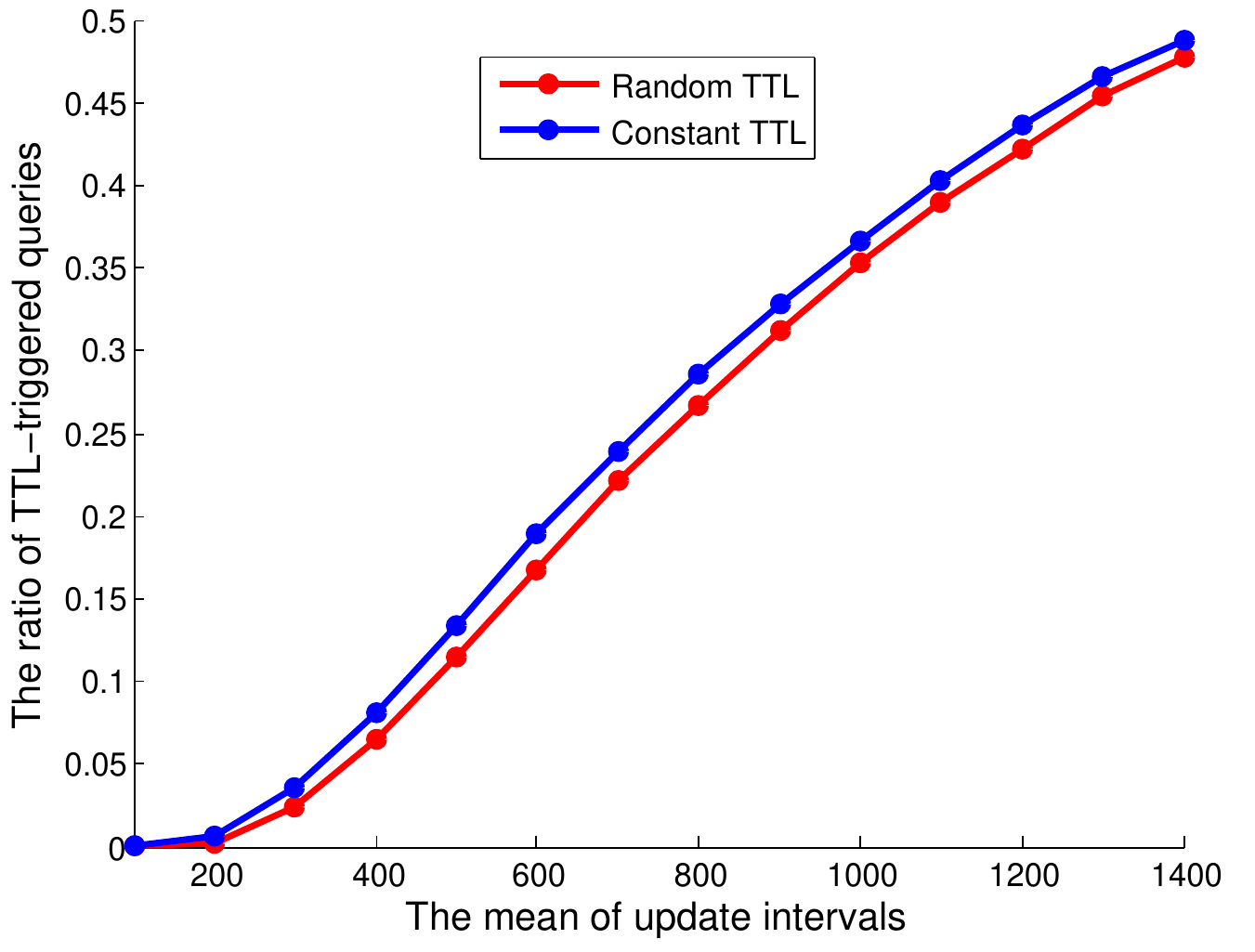}
%\vspace{-1em}
%\begin{flushleft}
%\begin{center}
%\footnotesize{Fig. 1. NSEC record chain}
%{\leftskip=0pt \rightskip=0pt plus 0cm
%Fig. 3. Ratio of TTL-triggered queries vs authoritative update intervals.
%}
\caption{Ratio of TTL-triggered queries vs authoritative update intervals.}
\vspace{-2em}
%\end{flushleft}
%\end{center}
\end{minipage}
\end{figure}

The window extension mechanism will dramatically diminish the success rate of cache poisoning in a given time because one round of cache poisoning attempt with a constant success rate of cache poisoning just lasts much longer. The cached validating records suppress the attacker from initiating a new round of cache poisoning attempt immediately after the old round proves a failure. A new window starts whenever the validating records expires from the cache. So the length of window is at least the TTL of the validating records. Furthermore, the window may be prolonged to above a TTL if updates take place before the TTL expires. In such cases, the continuous elapse of TTL is interrupted by any update which renews the residual TTL to a full TTL. Then the cached validating records will still at least last a full TTL to expires its TTL. During this period, the residual TTL may be renewed again and again whenever a update occurs before it reaches zero. In general, the effects of window extension are better pronounced for a more frequent update.

\section{Model Checking Results}

\begin{table}[!htb]
\centering
\caption{Parameters and their settings.}\label{tab:1}
\begin{tabular}{p{7.5cm}p{1cm}}
%\hlinewd{0.5pt}
\toprule
Parameter & Setting\\
%\hlinewd{1pt}
\midrule
Number distinct IDs available & 65536\\
Number of ports used (ports less than 1024 are unavailable) & 64000 \\
Number of authoritative servers for a domain & 2.5\\
Window of opportunity & 0.02 s\\
Number of identical outstanding queries of a resolver & 20\\
Query sending rate from the target resolver to the authoritative servers & 100 qps\\
Query responding rate from the authoritative servers to the target resolver & 100 qps\\
Query sending rate from the attacker to the target resolver in order to create the outstanding requests	& 1000 qps\\
ToD	& 3\\
Bogus responding rate from the attacker to the target resolver	& 100\\
Life cycle of validating records & 10 hour\\
%\hline
\bottomrule
\end{tabular}
\end{table}

Probabilistic model checking is one of the most common used formal verification technique for the modelling and analysis of stochastic systems. 
%In probabilistic model checking, the construction and analysis of a probabilistic model explores all possible all possible states that can occur as well as all possible process scheduling.
PRISM \cite{PRISM} is an open-source probabilistic model checker, providing support for building and analyzing several types of probabilistic models.
%: discrete-time Markov chains (DTMCs), continuous-time Markov chains (CTMCs), Markov decision processes (MDPs), etc. plus extensions of these models with costs and rewards.
We model Kaminsky cache poisoning attack as a continuous-time Markov chain (CTMC) using PRISM. In modeling the attack, we assume that in each round of cache poisoning attempt, the queries originated from the attacker's client look up a random generated domain such that they will never hit the target resolver's cache. 
%Attackers are very likely to adopt this strategy in order to strengthen their attack efficacy.
We also assume that the IP addresses of the target domain's authoritative servers are always maintained in the cache of the target resolver. The assumption is reasonable because the TTL of authoritative servers' records are usually much longer than the duration of a cache poisoning attempt.

\subsection{Results of Query Load}

To investigate the effects of combination of TTL expiration and authoritative update on the intertime of DNSSEC queries, we generate a sequence of authoritative update events following a probabilistic distribution while setting the TTLs in the DNSSEC responses as constant and probabilistic values respectively. In one set of experiments, we let the TTLs of records in question be evenly distributed on the interval from 500s to 1500s. In the other set of experiments, the TTLs of records in question take a constant value as 1000s. For both sets of experiments, the intertime of authoritative updates follows exponential distribution with the parameter ranging from 100s to 1400s.

We use Monte Carlo method to estimate the mean of intertimes of DNSSEC queries. In each experiment, 100,000 times of authoritative updates are generated from an exponential distribution. A number of TTLs, taking either constant values or probabilistic values, are also produced to cover the same time span at the instances when the predecessor TTL expires or authoritative update takes place.

\begin{figure}[!t]
\centering
{\includegraphics[scale=0.31]{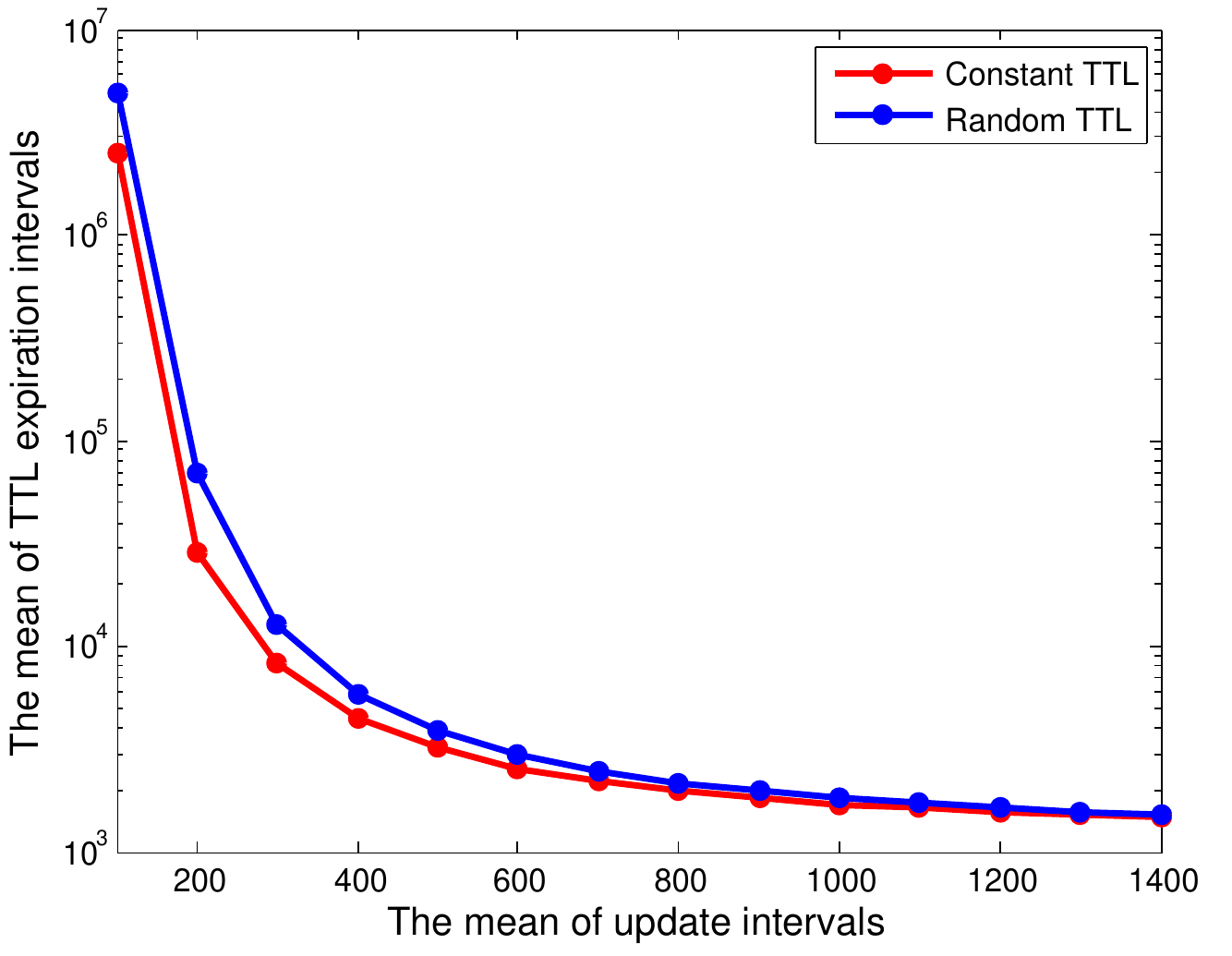}}
%\footnotesize{Fig. 1. NSEC record chain}
\caption{TTL expiration intervals vs authoritative update intervals.}
\vspace{-2em}
\end{figure}

Fig. 5 illustrates how DNSSEC query intervals change with authoritative update intervals. We can see that a very small authoritative update interval has almost the same DNSSEC query interval because TTL expiration rarely happens. But for a larger authoritative update interval, the limiting effect of TTL is better pronounced because a TTL has more chance of being smaller than an authoritative update interval thus more chance of expiration. Random TTLs, though have the same mean as constant TTLs, tend to cause a slightly larger DNSSEC query intervals and thereby a smaller DNSSEC query load on authoritative servers. The ratio of TTL-triggered queries is illustrated in Fig. 6.  We can see that the ratio of TTL-triggered queries grows as the mean of update intervals increases. But the authoritative update tends to pronounce more than TTL expiration on triggering DNSSEC queries even if they share the same mean interval. As shown in Fig. 7, when both update interval and TTL take a mean of 1000s,  TTL-triggered DNSSEC queries only account for about 36\% of the total. That can be explained by the fact that the event of authoritative update is independent of and never superceded by the event of TTL expiration while the even arrival of TTL expiration may be interrupted and restarted by an authoritative update.

It is obvious that DNSSEC query interval will be larger if authoritative update and TTL expiration are independent. So in order to examine the lower bound of DNSSEC query interval or the upper bound of DNSSEC query rate, we assume that authoritative update and TTL expiration are independent. Then the mean DNSSEC query interval can be written as
\begin{equation}
I_{overall}=\frac{I_{update}*I_{ttl}}{I_{update}+I_{ttl}}
\end{equation}
Where $I_{update}$  and $I_{ttl}$  represent the authoritative update interval and the TTL respectively.

As can be seen from Fig. 5 and Fig. 6, we can conclude that the maximum DNSSEC query rate of TDWN under intense cache poisoning attempts is of the same order as the minimum of the authoritative record update rate and the reciprocal of TTL.

\begin{figure}[!t]%{\captionstyle{normal}}
%\begin{minipage}[b]{0.5\linewidth}
\begin{minipage}[t]{0.48\linewidth}
\centering
\includegraphics[scale=0.31]{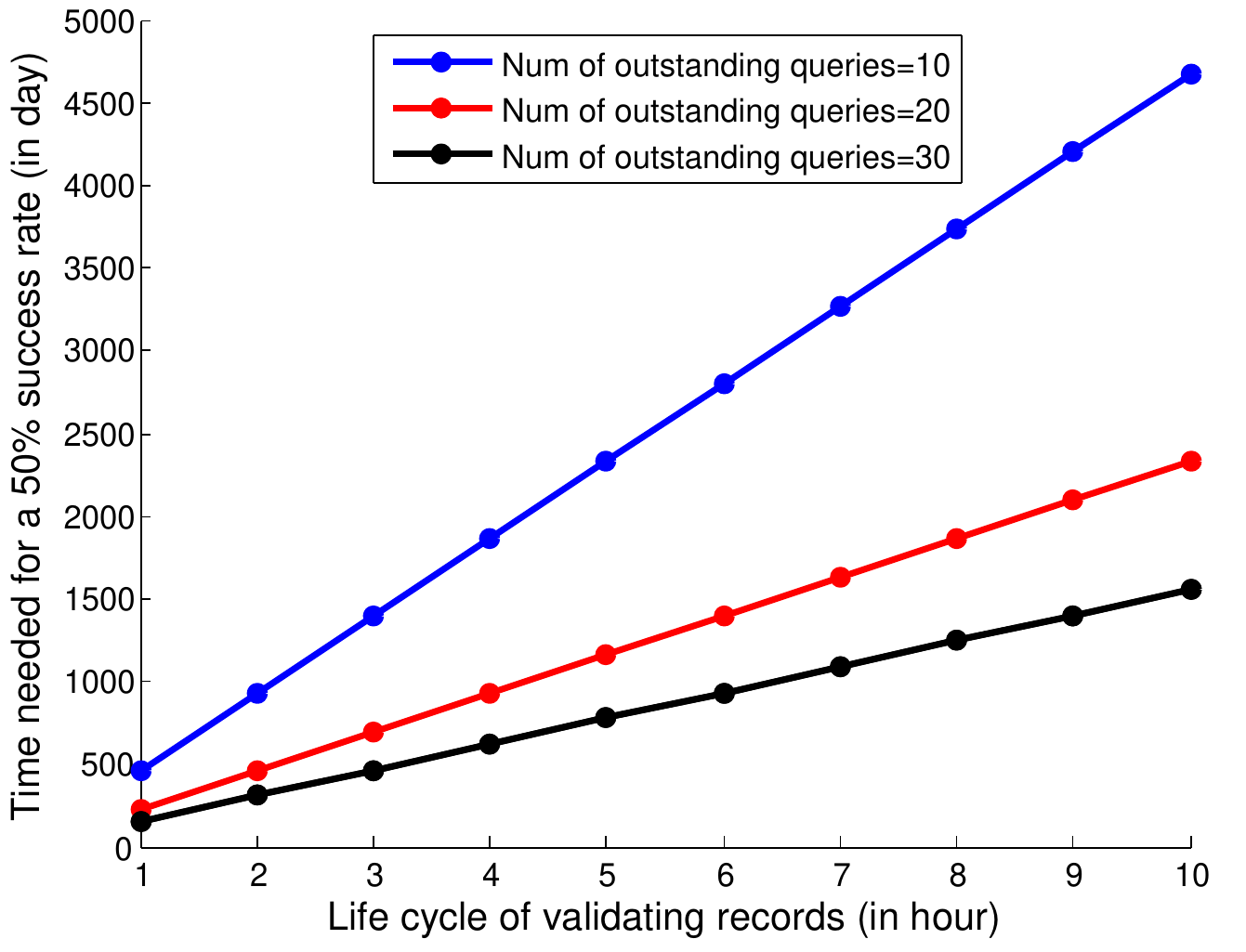}
\vspace{-1em}
%\captionstyle{normal}
\caption{Time needed for a 50\% success rate vs life cycle of validating records (ToD=3).}
%\begin{flushleft}
%\footnotesize{Fig. 1. NSEC record chain}
\vspace{-2em}
%\end{flushleft}
\end{minipage}
\hspace{0.3cm}
%\begin{minipage}[b]{0.5\linewidth}
\begin{minipage}[t]{0.48\linewidth}
\centering
\includegraphics[scale=0.31]{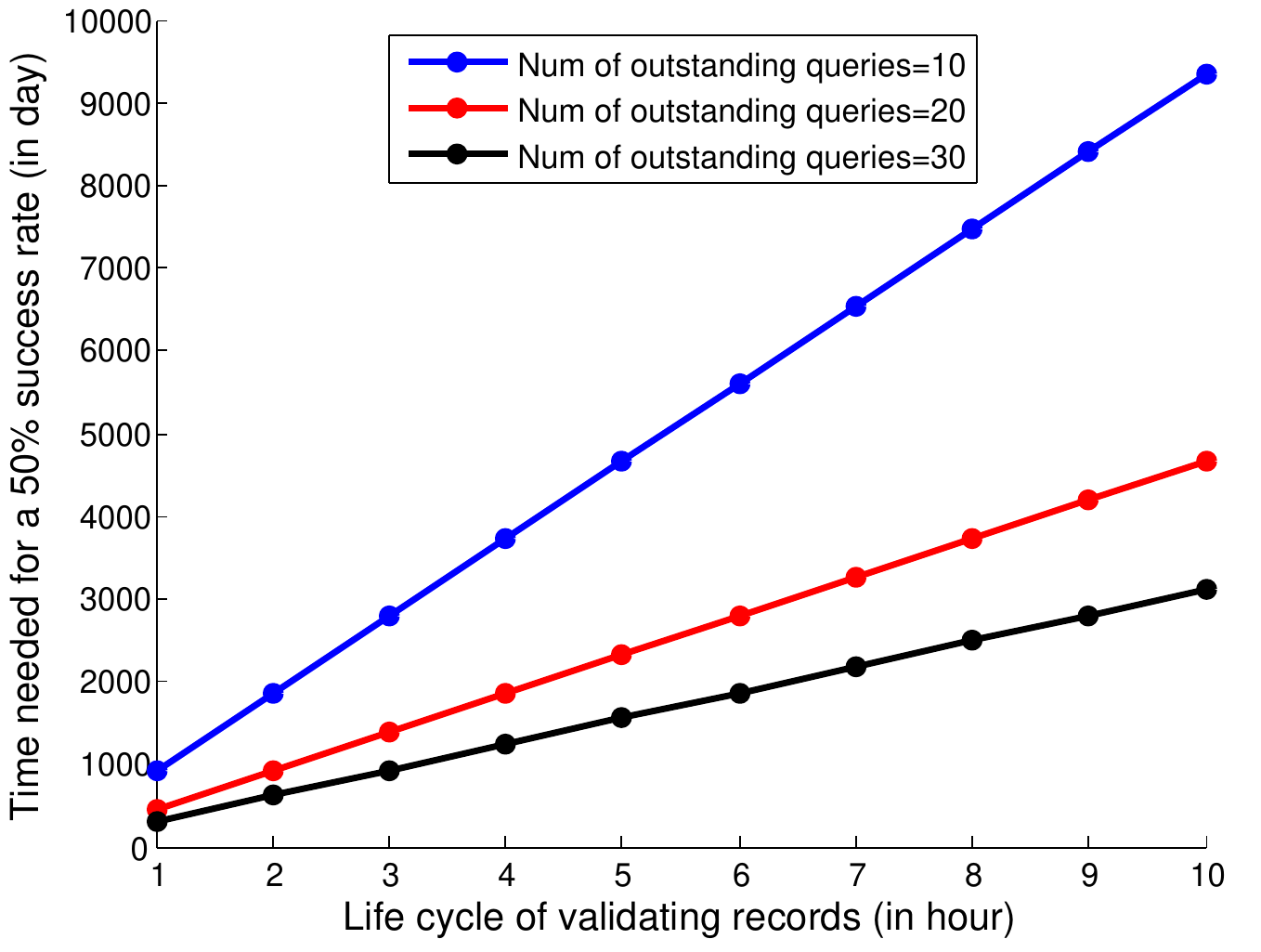}
%\vspace{-1em}
%\begin{flushleft}
%\begin{center}
%\footnotesize{Fig. 1. NSEC record chain}
%{\leftskip=0pt \rightskip=0pt plus 0cm
%Fig. 3. Ratio of TTL-triggered queries vs authoritative update intervals.
%}
\caption{Time needed for a 50\% success rate vs life cycle of validating records (ToD=2).}
\vspace{-2em}
%\end{flushleft}
%\end{center}
\end{minipage}
\end{figure}

\subsection{Results of Cache Poisoning Success Rate}

We configure the default values in Tab. 1 for the parameters in the model checking unless their values are otherwise stated.

First, we illustrate the time needed for a 50\% success rate under different life cycle of validating records in Fig. 8 (ToD=3).  We can see that the time cost of cache poisoning roughly grows linearly with the life cycle of validating records. For a life cycle above 10 hours, the time required for a 50\% success rate amounts to no less than 2 years. This is because the longer are the validating records  provided in cache to defend against cache poisoning attacks, the longer does an attacker have to wait to embark the next cache poisoning attempt (if the current attempt fails). As the TTLs of many authoritative records are set in the order of days or even weeks, it is very hard in practice to compromise them through cache poisoning attacks. Fig. 8 also shows creating more outstanding queries may dramatically decrease the difficulty of cache poisoning. Thus in the defense, the resolver should not allow excessive identical outstanding queries in order to prevent an unacceptable success rate of cache poisoning.

Second, we investigate the impacts of ToD on the success rate. In Fig. 9, the time needed for a 50\% success rate is shown when the ToD is lowered to 2. We can see that limiting ToD helps significantly to suppress the success rate of cache poisoning. Since ToD defines the maximum number of forgery responses (ToD-1) allowed without defense, a larger ToD means more chance of guessing attempts in a cache poisoning attempt thus a larger success rate. To ensure the efficacy of TDWN,  ToD should be set as a sound small value.

Third, we study how the cache poisoning success rate evolves over time. In Fig. 10, we can see that the success rate over time grows like a stair-step shape. In the curve, each step virtually represents a cache poisoning attempt in time and an accumulation of ToD-1 forgery responses in success rate. And the width of each stair-step is dominated by the life cycle of validating response. When ToD is three in Fig. 10, there are two forgery responses aggregated in a cache poisoning attempt to increase the overall success rate.

Fourth, how the setting of ToD impacts the cache poisoning success rate is studied. As illustrated in Fig. 11, the increase of ToD from 3 to 5 will lessen the defense of TWND against cache poisoning attacks. While the width of each stair-step stays the same as Fig. 10, the jump of each stair-step in the success rate is doubled. So the overall success rate grows much faster than Fig. 10.  This shows again that a large ToD may undermine the defense capability of TWND.

\section{Conclusions}

DNSSEC deployment suffers from its significant costs which in turn slow its progess. The resulting long transition to DNSSEC leaves a large name space still vulnerable to cache poisoning attacks. To speed up DNSSEC adoption and thereby narrow the window of transitional risks, a lightweight DNSSEC solution was proposed. The attack detection performed by recursive resolvers is employed to take up DNSSEC on demand rather than incessantly. The lightly used DNSSEC not only greatly lowers the DNSSEC overheads but also basically reserves the DNSSEC defense capability against cache poisoning attacks. 

\begin{figure}[!t]
%\begin{minipage}[b]{0.5\linewidth}
\begin{minipage}[t]{0.48\linewidth}
\centering
\includegraphics[scale=0.31]{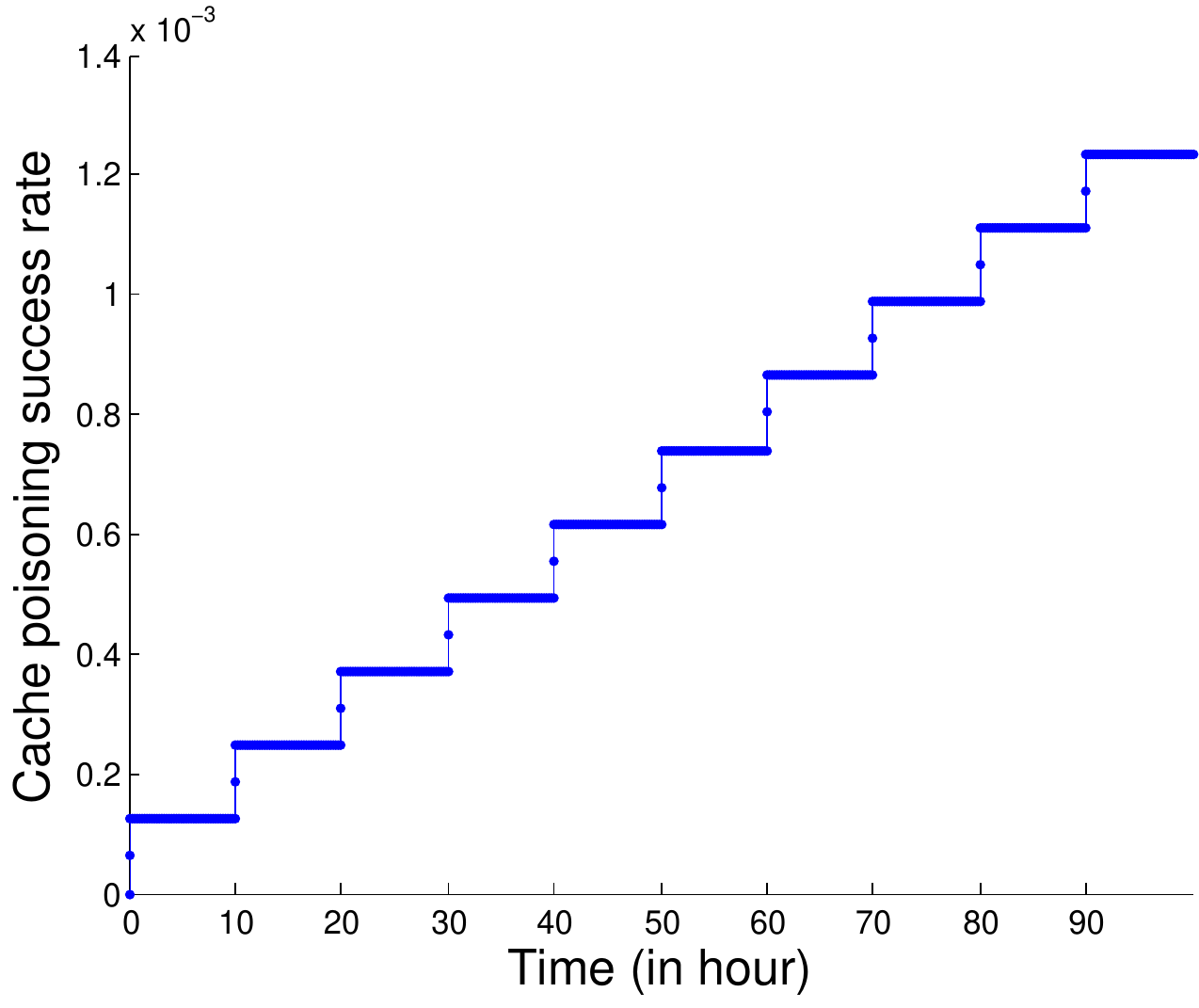}
\vspace{-0.5em}
%\begin{flushleft}
%\footnotesize{Fig. 1. NSEC record chain}
\caption{Cache poisoning success rate vs time (ToD=3).}
\vspace{-2em}
%\end{flushleft}
\end{minipage}
\hspace{0.3cm}
%\begin{minipage}[b]{0.5\linewidth}
\begin{minipage}[t]{0.48\linewidth}
\centering
\includegraphics[scale=0.31]{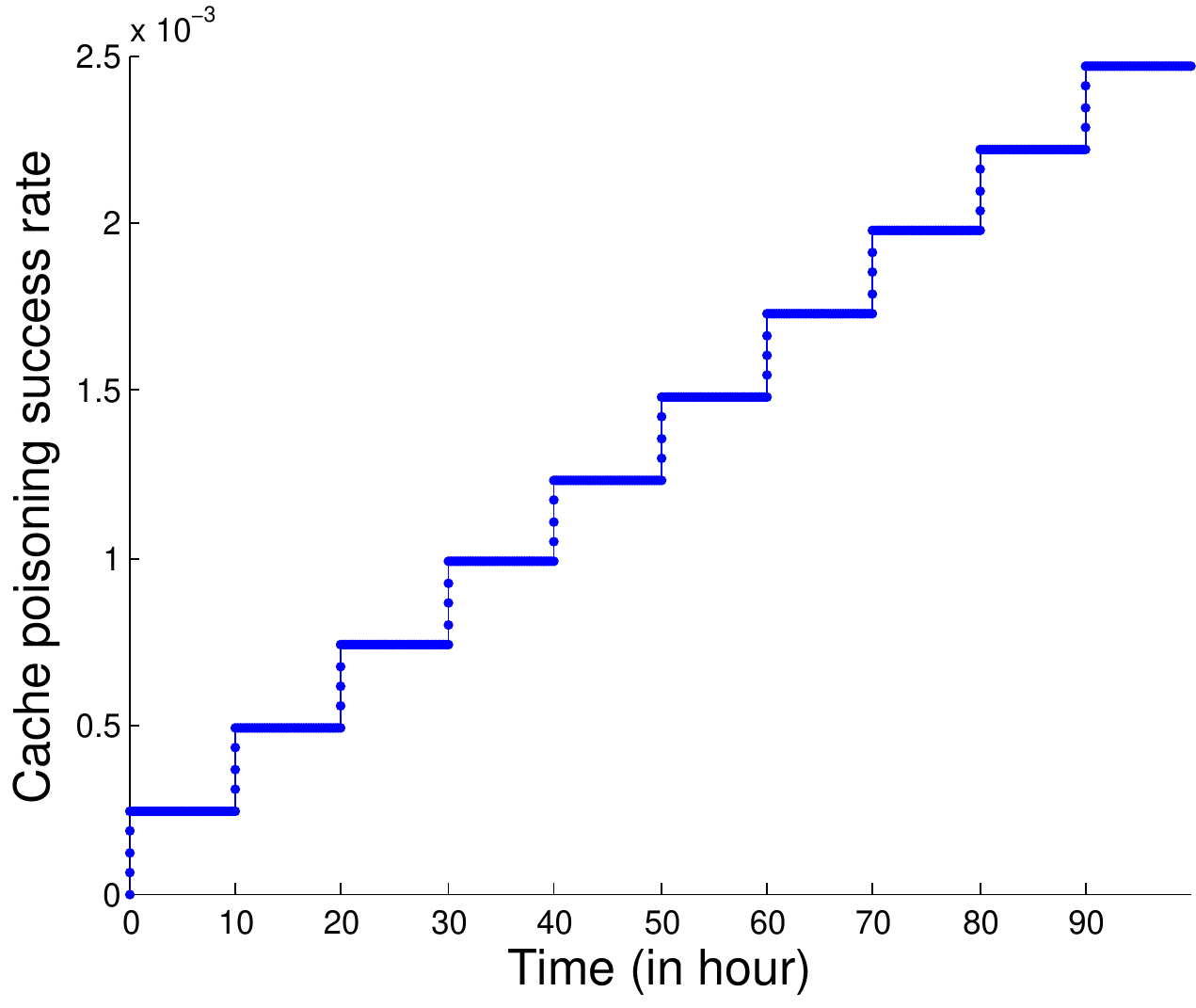}
\vspace{-0.5em}
%\begin{flushleft}
%\footnotesize{Fig. 1. NSEC record chain}
\caption{Cache poisoning success rate vs time (ToD=5).}
\vspace{-2em}
%\end{flushleft}
\end{minipage}
\end{figure}

\end{document}